\documentclass[preprint,showpacs,amsmath,amssymb]{revtex4}

\usepackage{mathrsfs}

\begin{document}


\title{Generalized Dirichlet normal ordering in open bosonic strings}


\author{Zhen-Bin Cao}
\email{caozhb04@lzu.cn}

\author{Yi-Shi Duan}

\affiliation{Institute of Theoretical Physics, Lanzhou University,
Lanzhou 730000, P. R. China}

\date{\today}


\begin{abstract}

Generally, open string boundary conditions play a nontrivial role in
string theory. For example, in the presence of an antisymmetric
tensor background field, they will lead the spacetime coordinates
noncommutative. In this paper, we mainly discuss how to build up a
generalized Dirichlet normal ordered product of open bosonic string
embedding operators that satisfies both the equations of motion and
the generalized Dirichlet boundary conditions at the quantum level
in the presence of an antisymmetric background field, as the
generalized Neumann case has already been discussed in the
literature. Further, we also give a brief check of the consistency
of the theory under the newly introduced normal ordering.


\end{abstract}

\maketitle

\smallskip

\section{Introduction}                                                          

Recent developments in string theory \cite{maldacena98, rs99-1,
rs99-2} suggest scenarios that our four-dimensional spacetime with
the standard model fields corresponds to a D3-brane
\cite{polchinski95} embedded in a larger manifold, on which the open
string endpoints attach. One important consequence of such models is
that, in the presence of an antisymmetric tensor background field,
the open string boundary conditions lead to the noncommutativity of
the spacetime coordinates on the branes. This `noncommutative
geometry' now has proven to play a key role in the dynamics of
D-branes, and is regarded to provide an important hint about the
nature of the spacetime at very small length scales \cite{chu99,
chu00, alekseev99, sieberg99}, which is one of the reasons for the
increasing interests in studying noncommutative quantum field
theories \cite{sieberg99, szabo03}. Furthermore, this also
illustrates the fact that the open string boundary conditions may
play a nontrivial role in string theory and in our four-dimensional
physics.

In the context of the quantum field theory, to preserve causality,
products of quantum fields must be defined at the same spacetime
points. But to do this, one also introduces divergences. The same
thing holds true in string theory, for example, products of the
string embedding operators on the same world-sheet points are in
general singular objects. This situation is well known and one can
remove the singular parts of the operator products by defining
proper normal ordered objects. Usually in string theory, the normal
ordered products of the conformal operators on the string
world-sheet are defined so as to satisfy the classical equations of
motion at the quantum level. Also for the open string case, as the
world-sheet has boundaries, they have to satisfy corresponding
boundary conditions. This has already been studied by Braga
\emph{etc}. In a recent paper \cite{braga05}, they defined a new
normal ordered product for open string embedding operators which
satisfy both the equations of motion and the boundary conditions
with a constant antisymmetric tensor background field present. But
what they considered was actually only the generalized Neumann case
in the world-volume of the D-branes, whose boundary conditions are a
mixture of the simple Neumann and Dirichlet ones. Meanwhile the two
boundary conditions can also mix in another way to form so-called
generalized Dirichlet boundary conditions in the directions
perpendicular to the D-branes which represent new information. So it
is quite necessary to study this generalized Dirichlet case.

In this paper, after a brief review of the newly defined Neumann
normal ordering in Ref. \cite{braga05} in Sec. \ref{Neumann-normal},
we will give a detailed discussion on a normal ordered operator
product which satisfies both the equations of motion and the
generalized Dirichlet boundary conditions in Sec.
\ref{Dirichlet-normal}. As the presence of the antisymmetric
background field, the discussion similar to that of the Neumann case
will not be explicit, and so we take another approach, namely we
consider the $d$-dimensional spacetime with some periodic
compactified dimensions, and take the radius of these dimensions
$R\rightarrow0$, which leads to corresponding $T$-dual noncompact
dimensions, leaving several D-branes at some places. Then we will
obtain the new Dirichlet boundary conditions and define a new
generalized Dirichlet normal ordered operator product in the
$T$-dual picture of the theory. We also check some further issues in
Sec. \ref{further-issues}, such as the central charge gets no
impact, and the spacetime coordinates of the string endpoints in the
directions parallel and perpendicular to the D-branes both become
noncommutative under these generalized Dirichlet and Neumann normal
orderings. Finally, the conclusion is given in Sec.
\ref{conclusion}.

\section{Review of Neumann normal ordering}
\label{Neumann-normal}

The generalized normal ordered operator product for the open bosonic
string which satisfies both the equations of motion and the
generalized Neumann boundary conditions in the D-brane world-volume
has recently been discussed in Ref. \cite{braga05} and further in
Ref. \cite{chakraborty06}. Here we give them a little detailed
review, as in which many results will be useful in the next section.
We generally use the same symbols as in Ref.
\cite{Polchinski98-string-book} in all this paper.

The classical action for a bosonic string in the presence of a
constant antisymmetric tensor background field $B_{\mu\nu}$ is
\begin{equation}\label{action-1}
S=\frac{1}{4\pi\alpha'}\int_\Sigma d^2\sigma(g^{ab}\eta_{\mu
  \nu}+i\epsilon^{ab}B_{\mu\nu})\partial_aX^\mu\partial_bX^\nu,
\end{equation}
where $X^\mu$ are the spacetime embedding coordinates of the bosonic
string, $\Sigma$ is the string world-sheet parameterized by
$\sigma^1=\sigma, \sigma^2=i\tau$ with the boundary (string
endpoints) for open string at $\sigma=0, \pi$, $g_{ab}$ is the
Euclidean world-sheet metric with signature $(+,+)$, $\epsilon^{ab}$
is the antisymmetric tensor with $\epsilon^{12}=1$, and
$\eta_{\mu\nu}=diag(-,+,\cdots,+)$ is the flat Minkowski spacetime
metric.

The variation of the action (\ref{action-1}) with respect to $X^\mu$
gives the equations of motion
\begin{equation}\label{eom-1}
(\partial^2_1+\partial^2_2)X^\mu=0,
\end{equation}
and the boundary conditions
\begin{equation}\label{bc-1}
(\eta_{\mu\nu}\partial_1+iB_{\mu\nu}\partial_2)X^\nu
 |_{\sigma=0,\pi}=0.
\end{equation}
Comparing to the original Neumann boundary conditions
$\partial_1X^\mu|_{\sigma=0,\pi}=0$ \cite{Polchinski98-string-book}
obtained with $B_{\mu\nu}=0$, Eqs. (\ref{bc-1}) mix both the Neumann
and Dirichlet boundary conditions, but we will simply call them the
generalized Neumann boundary conditions, to be consistent with the
discussion in the next section.

It is convenient to express the above equations in terms of complex
world-sheet coordinates. Define $w=\sigma^1+i\sigma^2,\;
\bar{w}=\sigma^1-i\sigma^2$, and further $z=-e^{-iw},\;
\bar{z}=-e^{i\bar{w}}$ \cite{Polchinski98-string-book}. As discussed
in Ref. \cite{braga05}, the classical action (\ref{action-1}) then
takes the form
\begin{equation}\label{action-2}
S=\frac{1}{2\pi\alpha'}\int_\Sigma d^2z(\eta_{\mu\nu}+B_{\mu\nu})
  \partial_zX^\mu\partial_{\bar{z}}X^\nu,
\end{equation}
and the classical equations of motion (\ref{eom-1}) and the
generalized Neumann boundary conditions (\ref{bc-1}) take the forms
\begin{eqnarray}
\partial_z\partial_{\bar{z}}X^\mu(z,\bar{z})\!&=&\!0,
 \label{eom-2} \\
\big[\eta_{\mu\nu}(\partial_z-\partial_{\bar{z}})-B_{\mu\nu}
(\partial_z+\partial_{\bar{z}})\big]X^\nu|_{z=\bar{z}}\!&=&\!0.
 \label{bc-2}
\end{eqnarray}

One can also study the properties of quantum operators by
considering the expectation values of the corresponding classical
objects. Define the value of an operator $\mathscr{F}$ by the path
integral
\begin{equation}
\big<\mathscr{F}[X]\big>=\int[dX]\exp(-S)\mathscr{F}[X].
\end{equation}
As discussed in Ref. \cite{Polchinski98-string-book}, by using the
fact that the path integral of a total derivative is zero,
\begin{equation}
0=\int[dX]\frac{\delta}{\delta X^\mu(z,\bar{z})}\exp(-S[X])
\end{equation}
will suggest that it holds the quantum versions of the equations of
motion (\ref{eom-2}) and the generalized Neumann boundary conditions
(\ref{bc-2}) (for more details, to see Ref. \cite{braga05})
\begin{eqnarray}
\partial_z\partial_{\bar{z}}\hat{X}^\mu(z,\bar{z})\!&=&\!0,
 \label{eom-3} \\
\big[\eta_{\mu\nu}(\partial_z-\partial_{\bar{z}})-B_{\mu\nu}
 (\partial_z+\partial_{\bar{z}})\big]\hat{X}^\nu|_{z=\bar{z}}
 \!&=&\!0,
 \label{bc-3}
\end{eqnarray}
where $\hat{X}^\mu$ now are the embedding operators of the string,
which means that these equations hold at the quantum level. Further,
products of operators at the same world-sheet points yield a
singular behavior. By calculating
\begin{equation}
0=\int[dX]\frac{\delta}{\delta X^\mu(z,\bar{z})}
  \big[\exp(-S[X])X^\nu(z',\bar{z}')\big],
\end{equation}
one finds the corresponding products of operators satisfy
\begin{eqnarray}
\frac{1}{\pi\alpha'}\partial_z\partial_{\bar{z}}
 \hat{X}^\mu(z,\bar{z})\hat{X}^\nu(z',\bar{z}')
 =-\eta^{\mu\nu}\delta^2(z-z',\bar{z}-\bar{z}'),
 \label{eom-4} \\
\big[\eta_{\mu\nu}(\partial_z-\partial_{\bar{z}})-B_{\mu\nu}
 (\partial_z+\partial_{\bar{z}})\big]\hat{X}^\nu(z,\bar{z})
 \hat{X}^\rho(z',\bar{z}')|_{z=\bar{z}}=0.
 \label{bc-4}
\end{eqnarray}
If defining a normal ordered product of two embedding operators in
the standard way \cite{Polchinski98-string-book}
\begin{equation}\label{Standard-normal}
:\hat{X}^\mu(z,\bar{z})\hat{X}^\nu(z',\bar{z}'):
 =\hat{X}^\mu(z,\bar{z})\hat{X}^\nu(z',\bar{z}')+\frac{\alpha'}
  {2}\eta^{\mu\nu}\ln|z-z'|^2,
\end{equation}
it will satisfy the equations of motion (\ref{eom-4}) but fail to
satisfy the generalized Neumann boundary conditions (\ref{bc-4}) at
the quantum level. To reconcile this problem, Braga \textit{etc}
recently introduced a new normal ordering \cite{braga05}
\begin{eqnarray}\label{Neumann-normal-2}
:\hat{X}^\mu(z,\bar{z})\!\!&&\!\!\hat{X}^\nu(z',\bar{z}'):\;
         =\hat{X}^\mu(z,\bar{z})\hat{X}^\nu(z',\bar{z}')
          +\frac{\alpha'}{2}\eta^{\mu\nu}\ln|z-z'|^2
          \nonumber  \\
   \!&&\!+\frac{\alpha'}{2}([\eta-B]^{-1}[\eta+B])^{\mu\nu}
          \ln(z-\bar{z}')+\frac{\alpha'}{2}([\eta-B][\eta+B]
          ^{-1})^{\mu\nu}\ln(\bar{z}-z').
\end{eqnarray}
It is easy to check that this normal ordered operator product
satisfies both the equations of motion (\ref{eom-4}) and the
generalized Neumann boundary conditions (\ref{bc-4}) at the quantum
level. To be consistent with the following discussion, here we name
this new normal ordered product as the second Neumann normal
ordering of the open bosonic string. When taking the antisymmetric
background field $B_{\mu\nu}=0$, it reduces to the familiar form
\begin{equation}\label{Neumann-normal-1}
:\hat{X}^\mu(z,\bar{z})\hat{X}^\nu(z',\bar{z}'):
 =\hat{X}^\mu(z,\bar{z})\hat{X}^\nu(z',\bar{z}')
  +\frac{\alpha'}{2}\eta^{\mu\nu}\ln|z-z'|^2
  +\frac{\alpha'}{2}\eta^{\mu\nu}\ln|z-\bar{z}'|^2,
\end{equation}
which is correspondingly called the first Neumann normal ordering.

\section{Dirichlet normal ordering}
\label{Dirichlet-normal}

To obtain the above generalized Neumann normal ordering, the autorhs
implicitly only considered things in the D-brane world-volume and
used the assumptions that the field $B=B_{\mu\nu}dX^\mu\wedge
dX^\nu$ and further the boundary conditions
$X^\alpha|_{\sigma=0,\pi} =x^\alpha_0$ (namely $\delta
X^\alpha|_{\sigma=0,\pi}=0$) \cite{chu99, chu00, jing05}, where now
we use the indices $\mu,\nu$ to denote the directions along which
the D-branes are expanded and the indices $\alpha,\beta$ to denote
the directions that are perpendicular to the D-branes. Here we note
that generally it is not quite proper to use the simple boundary
conditions $X^\alpha|_{\sigma=0,\pi} =x^\alpha_0$, as they only
allow flat and static D-branes, but which can actually be dynamical,
especially with nontrivial antisymmetric field components
$B_{\alpha\beta}$. So we need to generalize them in the directions
perpendicular to the D-branes. Also to make the discussion simple
and physically clear, we consider the case for the background field
$B=B_{\mu\nu}dX^\mu\wedge dX^\nu +B_{\alpha\beta}dX^\alpha\wedge
dX^\beta$.

Recall from the toroidal compactification and $T$-duality theories
\cite{Polchinski98-string-book} that when a spacetime dimension of a
string theory is periodically compactified $X\cong X+2\pi R$, the
field $X$ on the world-sheet will split into holomorphic and
antiholomorphic parts
\begin{equation}
X(z,\bar{z})=X_L(z)+X_R(\bar{z}).
\end{equation}
Then if taking the radius $R\rightarrow0$, it will lead to a new
theory with a noncompact dimension described by
\begin{equation}
X'(z,\bar{z})=X_L(z)-X_R(\bar{z}),
\end{equation}
with several D-branes leaving at some places, on which the open
string endpoints are fixed. Generally, these are the same theory,
one written in terms of $X$ and one in terms of $X'$. And the
equivalence is known as $T$-duality. One of the most important
properties of $T$-duality is that it interchanges the original
Neumann and Dirichlet boundary conditions,
\begin{equation}\label{Neumann-Dirichlet}
\partial_1X=-i\partial_2X',\quad \partial_2X=i\partial_1X'.
\end{equation}

Now consider some originally compactified dimensions $X^\alpha$ in
the whole spacetime. The equations of motion and the generalized
Neumann boundary conditions with nontrivial $B_{\alpha\beta}$
present are still expressed by (\ref{eom-1}) and (\ref{bc-1}) (only
by changing the indices $\mu,\nu$ to $\alpha,\beta$). But when
taking $R\rightarrow0$, in the $T$-dual picture, by considering
(\ref{Neumann-Dirichlet}), they generally change to the forms
\begin{eqnarray}
(\partial^2_1+\partial^2_2)X'^\alpha\!&=&\!0,
 \label{eom-6} \\
(\eta_{\alpha\beta}\partial_2-iB_{\alpha\beta}\partial_1)X'^\beta
 |_{\sigma=0,\pi}\!&=&\!0.
 \label{bc-6}
\end{eqnarray}
From these equations, we see that the equations of motion do not
change the forms, but the forms of the boundary conditions change.
Similar to the generalized Neumann boundary conditions (\ref{bc-1}),
these new ones can be called the generalized Dirichlet boundary
conditions, though they are also a mixture of the original Neumann
and Dirichlet ones. In these equations we use the prime to denote
the $T$-dual dimensions. But since all the following discussions
will be carried out in the $T$-dual noncompact dimensions, we will
drop the prime for simplicity. And here we can conclude the general
spacetime picture we will use in following and actually implicitly
used in the above section: we generally take $d$-dimensional
noncompact spacetime $X^M(M=0,\cdots,d-1)$, with several D$p$-branes
expanded in some directions $\alpha=p+1,\cdots,d-1$. (Also we should
specify that all the calculations in the above section were done in
the world-volume of the D-branes only).

To express the above equations (\ref{eom-7}) and (\ref{bc-7}) in
terms of complex world-sheet coordinates, it gives that
\begin{eqnarray}
\partial_z\partial_{\bar{z}}X^\alpha\!&=&\!0,
 \label{eom-7} \\
\big[\eta_{\alpha\beta}(\partial_z+\partial_{\bar{z}})-B_{\alpha\beta}
 (\partial_z-\partial_{\bar{z}})\big]X^\beta|_{z=\bar{z}}
 \!&=&\!0.
 \label{bc-7}
\end{eqnarray}
A similar path integral method show that these equations of motion
(\ref{eom-7}) and generalized Dirichlet boundary conditions
(\ref{bc-7}) also have quantum versions, namely the corresponding
operators satisfy
\begin{eqnarray}
\partial_z\partial_{\bar{z}}\hat{X}^\alpha(z,\bar{z})\!&=&\!0,
 \label{eom-8} \\
\big[\eta_{\alpha\beta}(\partial_z+\partial_{\bar{z}})-B_{\alpha\beta}
 (\partial_z-\partial_{\bar{z}})\big]\hat{X}^\beta|_{z=\bar{z}}
 \!&=&\!0,
 \label{bc-8}
\end{eqnarray}
at the quantum level. Further, the products of operators at the same
world-sheet points also yield a singular behavior,
\begin{eqnarray}
\frac{1}{\pi\alpha'}\partial_z\partial_{\bar{z}}
 \hat{X}^\alpha(z,\bar{z})\hat{X}^\beta(z',\bar{z}')
 =-\eta^{\alpha\beta}\delta^2(z-z',\bar{z}-\bar{z}'),
 \label{eom-9} \\
\big[\eta_{\alpha\beta}(\partial_z+\partial_{\bar{z}})-B_{\alpha\beta}
 (\partial_z-\partial_{\bar{z}})\big]\hat{X}^\beta(z,\bar{z})
 \hat{X}^\rho(z',\bar{z}')|_{z=\bar{z}}=0.
 \label{bc-9}
\end{eqnarray}
Again the standard normal ordering (\ref{Standard-normal}) satisfies
the equations of motion (\ref{eom-9}) but fails to satisfy the
generalized Dirichlet boundary conditions (\ref{bc-9}) at the
quantum level. And to reconcile this problem, similar to the second
Neumann normal ordering (\ref{Neumann-normal-2}), we define a new
normal ordering as
\begin{eqnarray}\label{Dirichlet-normal-2}
:\hat{X}^\alpha(z,\bar{z})\!\!&&\!\!\hat{X}^\beta(z',\bar{z}'):\;
         =\hat{X}^\alpha(z,\bar{z})\hat{X}^\beta(z',\bar{z}')+\frac
          {\alpha'}{2}\eta^{\alpha\beta}\ln|z-z'|^2   \nonumber  \\
   \!&&\!-\frac{\alpha'}{2}([\eta-B]^{-1}[\eta+B])^{\alpha\beta}\ln
          (z-\bar{z}')-\frac{\alpha'}{2}([\eta-B][\eta+B]^{-1}
          )^{\alpha\beta}\ln(\bar{z}-z'),
\end{eqnarray}
which satisfies both the equations of motion (\ref{eom-9}) and the
generalized Dirichlet boundary conditions (\ref{bc-9}) at the
quantum level. Comparing to the second Neumann normal ordering
(\ref{Neumann-normal-2}), this new one should be named as the second
Dirichlet normal ordering of the open bosonic string. When taking
the antisymmetric background field $B_{\alpha\beta}=0$, it reduces
to another familiar form
\begin{equation}\label{Dirichlet-normal-1}
:\hat{X}^\alpha(z,\bar{z})\hat{X}^\beta(z',\bar{z}'):
 =\hat{X}^\alpha(z,\bar{z})\hat{X}^\beta(z',\bar{z}')
  +\frac{\alpha'}{2}\eta^{\alpha\beta}\ln|z-z'|^2
  -\frac{\alpha'}{2}\eta^{\alpha\beta}\ln|z-\bar{z}'|^2,
\end{equation}
which is correspondingly called the first Dirichlet normal ordering.

It is interesting to notice that the above second Neumann
(\ref{Neumann-normal-2}) and Dirichlet (\ref{Dirichlet-normal-2})
normal orderings are quite likely, only with the signatures of the
additional terms being opposite. These additional terms can be
understood as an `image' charge contribution as in electrostatics,
but with different signs. For the second Neumann normal ordering
(\ref{Neumann-normal-2}), the image charge takes the same signature
as the original charge, and for the second Dirichlet normal ordering
(\ref{Dirichlet-normal-2}), the two charges take the opposite
signatures. This appears especially clear for the antisymmetric
field $B=0$ case (\ref{Neumann-normal-1}) and
(\ref{Dirichlet-normal-1}), which is already known. Further
comparing Eqs. (\ref{Neumann-normal-2}) and
(\ref{Dirichlet-normal-2}) to Eqs. (\ref{Neumann-normal-1}) and
(\ref{Dirichlet-normal-1}), it also suggests a physical
manifestation of the antisymmetric background field $B$ here: it
reflects the affection of one direction by the other ones. In the
case discussed here, as we have supposed $B_{\mu\alpha}=0$, a
direction perpendicular to the D-branes (or to say, an originally
compactified dimension) can only be affected by other perpendicular
directions through the second Dirichlet normal ordering, and the
same thing holds in the D-brane world-volumes (or to say, the
originally noncompact dimensions) through the second Neumann normal
ordering. In the more general case with nontrivial $B_{\mu\alpha}$
components, the two different sets of directions can also affect
each other through the two gneralized normal orderings.

\section{Further issues}\label{further-issues}

Now that we have obtained the generalized Neumann and Dirichlet
normal orderings of the open bosonic string in the presence of a
constant antisymmetric background field, for any arbitrary
functional operators $\mathscr{F}[X]$ and $\mathscr{G}[X]$, their
OPE is generalized to
\begin{eqnarray}
:\mathscr{F}::\mathscr{G}:\!&=&\!\exp\Big(-\frac{\alpha'}{2}\int
          d^2z_1d^2z_2\big[\eta^{\mu\nu}\ln|z_1-z_2|^2+\big([\eta
          -B]^{-1}[\eta+B]\big)^{\mu\nu}\ln(z_1-\bar{z}_2)
          \nonumber  \\
   \!&&\!+\big([\eta-B][\eta+B]^{-1}\big)^{\mu\nu}\ln(z_1-\bar{z}
          _2)\big]\frac{\delta}{\delta X^\mu_F(z_1,\bar{z}_1)}
          \frac{\delta}{\delta X^\nu_G(z_2,\bar{z}_2)}\Big)
          :\mathscr{F}\mathscr{G}:
\end{eqnarray}
in the directions along which the D-branes expand by using the
second Neumann normal ordering (\ref{Neumann-normal-2}), and to
\begin{eqnarray}
:\mathscr{F}::\mathscr{G}:\!&=&\!\exp\Big(-\frac{\alpha'}{2}\int
          d^2z_1d^2z_2\big[\eta^{\alpha\beta}\ln|z_1-z_2|^2-\big
          ([\eta-B]^{-1}[\eta+B]\big)^{\alpha\beta}\ln(z_1-\bar
          {z}_2)  \nonumber  \\
   \!&&\!-\big([\eta-B][\eta+B]^{-1}\big)^{\alpha\beta}\ln(z_1-
          \bar{z}_2)\big]\frac{\delta}{\delta X^\alpha_F(z_1,\bar
          {z}_1)}\frac{\delta}{\delta X^\beta_G(z_2,\bar{z}_2)}
          \Big):\mathscr{F}\mathscr{G}:
\end{eqnarray}
in the directions perpendicular to the D-branes by using the second
Dirichlet normal ordering (\ref{Dirichlet-normal-2}), where the
functional derivatives act only on the fields in $\mathscr{F}$ or
$\mathscr{G}$ respectively. And their complete OPE form is the sum
of these two forms.

Then it is important to check whether the theory is still consistent
under these changes. The first issue is the central charge, which is
a purely quantum effect and takes a central role in deciding the
critical dimensions of the string theory. Actually the check is
quite simple. The world-sheet energy-momentum tensor is
\begin{equation}
T(z)=-\frac{1}{\alpha'}:\partial_zX^M(z)\partial_zX_M(z):,
\end{equation}
where $M$ sums over $0,\cdots,d-1$, and we neglect the hats on the
embedding operators. Then one sees that as
$\partial_z\bar{z}=\partial_{\bar{z}}z=0$, it gives
\begin{equation}
\partial_z\ln(\bar{z}-z')=\partial_{\bar{z}}\ln(z-\bar{z}')=
\partial_{z'}\ln(z-\bar{z}')=\partial_{\bar{z}'}\ln(\bar{z}-z')=0.
\end{equation}
In the directions parallel to the D-branes, by using the second
Neumann normal ordering (\ref{Neumann-normal-2}), a direct
calculation yields
\begin{eqnarray}\label{TT-OPE-noncompact}
:\partial_zX^\mu(z)\partial_zX_\mu(z)::\partial_{z'}X^\nu(z')
          \partial_{z'}X_\nu(z'):
   \!&\sim&\!\frac{(p+1)\alpha'^2}{2(z-z')^4}-\frac{2\alpha'}
          {(z-z')^2}:\partial_{z'}X^\mu(z')\partial_{z'}X_\mu
          (z'):      \nonumber \\
   \!&&\!-\frac{2\alpha'}{z-z'}:\partial^2_{z'}X^\mu(z')\partial
          _{z'}X_\mu(z'):,
\end{eqnarray}
and in the directions perpendicular to the D-branes, by using the
second Dirichlet normal ordering (\ref{Dirichlet-normal-2}),
\begin{eqnarray}\label{TT-OPE-compact}
:\partial_zX^\alpha(z)\partial_zX_\alpha(z)::\partial_{z'}X^
          \beta(z')\partial_{z'}X_\beta(z'):
   \!&\sim&\!\frac{(d-1-p)\alpha'^2}{2(z-z')^4}-\frac{2\alpha'}
          {(z-z')^2}:\partial_{z'}X^\alpha(z')\partial_{z'}X_
          \alpha(z'):      \nonumber \\
   \!&&\!-\frac{2\alpha'}{z-z'}:\partial^2_{z'}X^\alpha(z')
          \partial_{z'}X_\alpha(z'):,
\end{eqnarray}
where $\sim$ means `equal up to nonsingular terms', and $p$ counts
the number of spatial dimensions of the D-branes, which suggests
that in (\ref{TT-OPE-noncompact}) the indices $\mu,\nu$ sum over
only the dimensions parallel to the D-branes and in
(\ref{TT-OPE-compact}) $\alpha,\beta$ over the dimensions
perpendicular to the D-branes. Combining these two relations implies
that
\begin{equation}
T(z)T(z')\sim\frac{d}{2(z-z')^4}+\frac{2}{(z-z')^2}T(z')
+\frac{1}{z-z'}\partial_{z'}T(z'),
\end{equation}
which is the same as the standard $TT$ OPE obtained by using the
original normal ordering (\ref{Standard-normal}) with the
antisymmetric field $B=0$. Here it should be noted that only by
considering both the second Neumann and Dirichlet normal orderings
in the two respect sets of directions can one get a complete and
correct result (and so it is not quite correct as did in Ref.
\cite{chakraborty06}, for they only considered the case in the
D-brane world-volume). The same relation holds for the
$\tilde{T}\tilde{T}$ OPE. Therefore, we get the results that the
Virasoro algebra remains unchanged and the two new generalized
normal orderings have no impact on the central charge.

Next we check the normal ordered commutators of the embedding
operators of the open bosonic string. For the second Neumann normal
ordering (\ref{Neumann-normal-2}), it was discussed in Ref.
\cite{braga05}, with the result in terms of the $\sigma^1,\sigma^2$
coordinates that
\begin{equation}\label{normal-commutator}
:[X^\mu(\sigma^1,\sigma^2),X^\nu(\sigma^1,\sigma'^2)]:
=[X^\mu(\sigma^1,\sigma^2),X^\nu(\sigma^1,\sigma'^2)],
\end{equation}
and for the second Dirichlet normal ordering
(\ref{Dirichlet-normal-2}), a totally similar calculation shows that
the result (\ref{normal-commutator}) still holds (only by changing
$\mu,\nu$ to $\alpha,\beta$), which together suggest that the
commutators do not get any extra contributions from these two
generalized normal ordering prescriptions. Then the equal time
commutators, expressed by identifying $\tau=\tau'$, i.e.
$\sigma^2=\sigma'^2$ and setting $\sigma^1=\sigma,
\sigma'^1=\sigma'$, are given
\begin{equation}
\big[X^\mu(\tau,\sigma),X^\nu(\tau,\sigma')\big]
   =2i\alpha'\big(M^{-1}B\big)^{\mu\nu}\Big[\sigma+
   \sigma'-\pi+\sum_{m\neq0}\frac{1}{m}\sin m(\sigma
   +\sigma')\Big],
\end{equation}
in the D-brane world-volume \cite{chakraborty06} and further a
similar calculation suggests
\begin{equation}
\big[X^\alpha(\tau,\sigma),X^\beta(\tau,\sigma')\big]
   =-2i\alpha'\big(M^{-1}B\big)^{\alpha\beta}\Big[\sigma+
   \sigma'-\pi+\sum_{m\neq0}\frac{1}{m}\sin m(\sigma
   +\sigma')\Big].
\end{equation}
in the directions perpendicular to the D-branes. As the infinite
series give
\[
\sum_{m\neq0}\frac{1}{m}\sin m(\sigma+\sigma')=\pi-(\sigma+\sigma')
\]
for $\sigma,\sigma'\in(0,\pi)$ and on the boundaries of the string
world-sheet
\[
\sum_{m\neq0}\frac{1}{m}\sin m(\sigma+\sigma')
|_{\sigma,\sigma'=0,\pi}=0,
\]
these commutation relations show that the coordinates of string
endpoints in the directions both parallel and perpendicular to the
D-branes become noncommutative, while in the internal of the open
string world-sheet, they are still commutative.

One can check other issues, such as the mode expansions of the
bosonic string \cite{chakraborty06}, and generally one finds that
though the spectrum is shifted in the presence of the antisymmetric
field, the theory is still consistent and the world-sheet is still a
CFT, a good string background.

\section{Conclusion}\label{conclusion}

Generally the boundary conditions for the open string take an
important role in many aspects of the string theory, such as the
quantization of the open string, the noncommutativity of the
spacetime embedding coordinates, etc. So they have been detailed
discussed in the literature, especially with a newly defined normal
ordered operator product, in the presence of a constant
antisymmetric background field. But meanwhile most discussions only
focused on the generalized Neumann case in the D-brane world-volume.
And so in this paper, after giving a brief review of the Neumann
case, we discussed the generalized Dirichlet case in the directions
perpendicular to the D-branes, by using a quite technical approach
that we take the radius of some periodic compactified dimensions to
zero and obtain the generalized Dirichlet boundary conditions in the
$T$-dual picture. Then to satisfy both the equations of motion and
the generalized Dirichlet boundary conditions, we defined a new
normal ordered product of the embedding operators. Finally we
discussed that both the generalized Neumann and Dirichlet normal
orderings, which are quite likely, though lead the spacetime
coordinates of the string endpoints in the directions both parallel
and perpendicular to the D-branes noncommutative, but have no impact
on the central charge, and retain the consistency of the theory.

\smallskip
\noindent {\bf Acknowledgments}\,                                               
This work was supported by the National Natural Science Foundation
and the Doctor Education Fund of Educational Department of the
People’s Republic of China.


\end{document}